\documentclass[12pt,preprint]{aastex}
\shortauthors{Li \& Smith}
\shorttitle{Multi-seeded clustering star formation in the RMC}

\received{2004 May 7}
\begin{document}

\title{Discovery of multi-seeded multi-mode formation of embedded
clusters in the Rosette Molecular Complex}
\author{Jin Zeng Li$^{1,2}$ \& Michael D. Smith$^{2}$}
\affil{$^{1}$National Astronomical Observatories, Chinese Academy of 
           Sciences, Beijing 100012, China (E-mail: ljz@bao.ac.cn) \\
	$^{2}$Armagh Observatory, College Hill, Armagh BT61 9DG, N. Ireland, UK \\}

\begin{abstract}
An investigation based on data from the spatially complete 2MASS Survey reveals 
that a remarkable burst of clustered star formation is taking place throughout 
the south-east quadrant of the Rosette Molecular Cloud. Compact clusters are 
forming in a multi-seeded mode, in parallel and at various places. In addition,
sparse aggregates of embedded young stars are extensively distributed.  
In this study, we report the primary results and implications for high-mass and 
clustered star formation in giant molecular clouds. In particular, we incorporate 
for the first time the birth of medium to low-mass stars
into the scenario of sequential formation of OB clusters.
Following the emergence of the young OB cluster NGC 2244, a variety 
of manifestations of forming clusters of medium to high mass 
appear in the vicinity of the swept-up layer of the H{\small II} region 
as well as further into the molecular cloud. 
The embedded clusters appear to form in a structured manner, which suggests they follow 
tracks laid out by the decay of macroturbulence. We address the possible origins
of the turbulence. This leads us to propose a
tree model to interpret the neat spatial distribution of
clusters within a large section of the Rosette complex. Prominent new generation OB 
clusters are identified at the root of the tree pattern.

\end{abstract}


\section{Introduction}

Giant Molecular Clouds (GMCs) are the primary sites of star birth in the Galaxy. 
In particular, they appear essential for the formation of massive stars in OB 
associations.
The Rosette Molecular Cloud (RMC), residing in the constellation of Monoceros, 
is situated in the Perseus arm of the Galaxy. It has an extent of around 
100\,pc (Dorland \& Montmerle 1987) and contains $\sim$~10$^{5}$~M$\odot$ of 
gas \& dust (Blitz \& Thaddeus 1980). It is among the most massive GMCs in 
the Galaxy and is associated with one of the most outstanding regions of massive 
star formation.  

The Rosette Nebula is located at the north-west tip of the GMC. It is a spectacular 
H{\small II} region of $\sim$~30\,pc in scale, famous for harboring one
of the youngest ($\sim$ 3\,$\times$\,10$^6$ yrs, Ogura \& Ishida 1981) open clusters, 
NGC~2244, and is itself being excavated by the strong stellar winds from the 
immersed OB stars at the center. The favorable orientation of the Rosette Nebula, 
perpendicular to the line of sight, and its prominent interaction with the ambient 
molecular cloud, provides an excellent opportunity to test the theory of sequential
formation of OB clusters in GMCs (Elmegreen \& Lada 1977; Lada 1987).

NGC~2244 is, on the other hand, located roughly at the center of the Mon~OB2 association, 
which has a physical scale of up to 280\,pc (Blaauw 1964; Singh \& Naranan 1979).
It represents the youngest subgroup of the OB association and extends $\sim$~20\,pc.
Another two subgroups of OB stars were identified and ages of the order of 
$\sim$5 x 10$^{6}$ yrs and a few 10$^{7}$ yrs suggested (Turner 1976).
One of their members could have exploded into a type II supernova and formed
a supernova remnant (SNR), the Monoceros Loop (Davies, 1963; Aharonian et al. 2004). However, the SNR has a
dynamical age of only 3-15 x 10$^{4}$ yrs, and the shock front has probably 
just reached the edge of the Rosette Nebula and is, therefore, unlikely to be responsible for any 
on-going star formation activity in the RMC (Blitz \& Thaddeus 1980; Townsley et al. 2003).
Nevertheless, the overall picture is that GMCs in the galactic disk
are transitory objects, subject to sequential cluster formation. 

The RMC is highly clumped, as has been established through various observations from
mid-infrared to radio wavelengths (Cox, Deharveng \&
Lenne 1990; Blitz \& Stark 1986; Williams, Blitz \& Stark 1995; Shipman \& Carey 1996; 
Kuchar \& Bania 1993;
Celnik 1985). It could be producing new generations of young stars and embedded 
clusters in its densest rim (Cox et al. 1990; Block, Geballe \& Dyson 1993; Carpenter, Snell \& 
Schloerb et al. 1993; Hanson, Geballe \& Conti et al. 1993; Phelps 1994). 
Patches of intensive X-ray emission spatially associated with the dense clumps, 
which is indicative of active star formation, were detected by
Gregorio-Hetem, Montmerle and Casanova et al. (1998).  

Phelps \& Lada (1997) 
have spotted, via visual inspection of near-infrared images, 7 aggregates of 
infrared stars, each associated with an IRAS point source. This demonstrated 
that star formation is continuing in this area in a distributed manner and different star
formation mechanisms may be at work. Unfortunately, a complete, unbiased census of star formation 
over the entire spatial extent of the RMC is not yet available. However, the 2MASS
JHKs data permit us to identify and analyze considerably more embedded objects.
A detailed, in-depth study will help to elucidate to what
extent star or cluster formation is proceeding and provide clues as to what mechanisms are governing
the star birth processes across this exceptional area. 
This Letter explores a panorama of clustered star formation in the RMC. 
Detailed studies of each well-defined region of cluster formation
will be presented in a forthcoming series. In addition, the first infrared study of 
the young massive cluster, NGC 2244, resolved into a combination of a 'twin' cluster and
a relic arc of slightly reddened young stars, will be discussed elsewhere
(Li 2005).

\section{Data Acquisition and Analysis}

Archived data from the 2MASS Point Source Catalog (PSC) and IRAS Sky Survey 
Atlases (ISSA) were retrieved via IRSA (http://irsa.ipac.caltech.edu/). IRAS
HIRES images of the RMC were subsequently obtained by the email requesting system running
at the IPAC data center. The following
sample selection criteria were employed to guarantee the reliability of 
the 2MASS data in use. (1) Each
source extracted from the 2MASS PSC must have a certain detection in all three 
of the J, H, and Ks bands.  (2) We restrict the Ks band signal to noise 
ratio to values above 15 to tightly constraint field stars in the control field to the 
main sequence loci on the (J-H) to (H-Ks) diagram (Li \& Smith 2005a). 
However, the flux and resolution 
limited 2MASS survey, and our stringent selection criteria, will in combination result
in drawbacks.  Firstly, the number of proto-OB stars that are deeply embedded can be 
underestimated. This is likely to occur in the densest regions with the highest extinction. 
Secondly, this work is limited to the study of comparatively high mass cloud members. 
With a 90$\%$ completeness limit to $\sim$~0.8~M$\odot$ objects, many disk stars with masses 
above this limit can be trimmed off.

\section{Results and Discussion}

Even in a single near-infrared band, two separate regions of
significant stellar density enhancement are readily identified.
One of these stellar concentrations coincides with the young open cluster
NGC 2244 and the other corresponds spatially to the densest rim of the RMC.
The latter has no known optical counterparts and is therefore likely to be
a recently formed embedded, compact cluster. 

However, when this distribution is plotted
as a function of H-Ks color, fine structure signifying large scale clustering
becomes evident throughout the south-east quadrant (Fig. 1). Positions of previously 
identified embedded clusters and
young massive stars are overplotted in Fig.~1. These are spatially associated with the
majority of the stellar concentrations in the plot, indicating the
young, embedded nature of the clustering of reddened sources. The candidate 
embedded clusters are further divided into four major groups A, B, C \& D, partially
according to their distinct concentrated appearance and their association with
dense clumps as outlined by the IRAS 100 $\mu$m emission
(Cox et al. 1990) and CO detections (Blitz \& Thaddeus 1980; Williams et al. 1995).
The majority of the candidate young stars are either congregated within the outlined
regions in Fig.~1 or appear as minor stellar aggregates not identified in this paper. 
Only a small fraction of the sources identified here with heavy line of sight extinction are 
found to be in a wide scattered distribution and are to be attributed to isolated formation.

Color-color diagrams of each of the four regions are presented in Fig. 2. About
one sixth to a fourth of the 2MASS sources in each region are found to be 
located to the right of the reddening band of normal field stars. These sources thus 
possess intrinsic color excesses originating from emission from circumstellar dust, commensurate
with their embedded nature.
A concentration of background giants along the left edge of the reddening
band directly above the giants track is apparent on the color-color diagram
for at least Region D, which covers the largest area in the RMC. This
indicates an inhomogeneous distribution of extinction for the cloud that ranges mainly
between 2--10~mag. In the densest clumps, however, visual extinction can even exceed 
25~mag. This confirms the very clumpy nature of the molecular
cloud and is consistent with a high clump to interclump density contrast (Schneider et al. 1998).

We attempt to eliminate foreground sources, which possess a typical color  of (H-Ks) $\sim$ 0.2 mag
based on a detailed investigation of the reference fields (Li \& Smith 2005a).
Apparent background giants concentrated within an elliptical area on the color-color
diagrams can also be excluded. The fraction of excess emission sources then remaining
in each region amounts to between one third and a half.
This provides convincing evidence that the majority of the reddened sources 
are embedded young stars still associated with their parental clouds. We thus
confirm the  unprecedented widespread presence of forming clusters embedded in the RMC. 

A close correspondence of the compact subclusters in each region with the
distribution of the most luminous IRAS sources (Cox et al. 1990) and the
densest clumps (Williams et al. 1995) is clearly shown in
Fig. 1. This suggests a `multi-seeded' origin of the embedded clusters in 
the fragmented interaction layer of the H{\small II} region and at various places deep 
into the molecular complex. The term multi-seeded is introduced to describe
a situation in which the formation of clusters in distinct clumps 
in a distributed manner 
is proceeding in parallel.
On the other hand, clustered star formation is taking place both 
in isolation in the swept-up shells and in a more structured mode further 
away from NGC~2244. The structured mode surprisingly continues down to the south-east boundary 
of the RMC.

The embedded nature of a source is thought to be fully substantiated if it meets the
following two criteria. These are (1) an intrinsic
color excess in the near infrared i.e. E(H-Ks) ${>}$ 0 and
(2) a (H-Ks) color of greater than 0.2. This again helps
to exclude possible contamination from the foreground. In this manner, some 30 
excessive emission sources were removed from our sample.
A clear density enhancement of embedded young stars toward
the outlined regions remains as evidence for a non-random distribution
of young stellar objects across the extincted areas. In addition, the 
surface density of excess emission sources ranges from an average of 0.55 pc$^{-2}$ 
over the entire extent of the RMC to $\sim$ 23 pc$^{-2}$ for the compact 
subclusters in Region C where the process of star formation is probably
still going on.

An investigation of the properties of each embedded cluster was carried out and 
the results are summarized in Table 1. The analysis included the construction of  
near infrared color-color and color-magnitude diagrams and the calculation
of the Ks Luminosity Function (KLF).
An example color-magnitude diagram for objects in the Region C cluster is presented in 
Fig. 3. Both components of the known binary system AFGL 961 are identified as 
candidate proto-O stars (for a detailed analysis, see Li \& Smith,
2005b) and are indicated with asterisks. The east component of AFGL 961,
presumably a single object, is shown to have a possible mass of up to $\sim$
130 M$\odot$. It is probably one of the most massive protostars forming in the complex.
A nearby progenitor O star, represented on the diagram with a solid square,
may prove to be a third component of the system
because of its close proximity to AFGL~961 within a projected distance of 0.2 pc
(Li \& Smith 2005b).

The clusters embedded in Regions A \& B are spatially coincident with the
fragmented arcs of the swept-up layer (Fig. 1 \& 4a) of the H{\small II} region
associated with NGC~2244, an emerging young open cluster known to have a turn-off 
age of $\sim$ 2~Myrs (Park \& Sung 2002). Based on a KLF study of the Regions A 
\& B clusters, an age estimate of $\sim$ 1~Myr is obtained by comparing to KLFs of embedded
clusters with known ages (Lada \& Lada 1995). At a distance of $\sim$ 20 pc away
from the center of NGC~2244, the age of the shell clusters in Regions A \& B
is consistent with probably their triggered nature by the shocked wave associated 
with the ionization front of 
the H{\small II} region, which propagates at a velocity of $\sim$ 20 km~s$^{-1}$ (Smith 1973).

No prominent age spread between the sub-clusters or aggregates in Regions C \& D
can be quantitatively addressed based on a study of the KLF. 
However, we suggest that the group of proto-O stars associated with AFGL 961, 
situated in the south-east of Region C (Fig. 1), is probably among the latest 
episodes of star and cluster formation. Due to their estimated high masses and their protostellar 
origin, this proto-O star system may have ages of less than a few 10$^{4}$ yrs 
(Hanson et al. 1993). At a distance of $\sim$ 15 pc from the interface with the 
Rosette Nebula, its location in the densest ridge of the RMC probably signifies 
the gestation of new generation OB clusters following the emergence
of NGC~2244.

\section{Structured clustering of star formation in the RMC}

As a comparison, we have derived the distributions of optical depth at 100 $\mu$m and 
color temperature based on the IRAS ISSA (Fig. 4a \& b, for an introduction of the methods used,  
see Li \& Chen 1996). The outlined regions A \& B are spatially coincident with
the swept-up layer of the Rosette Nebula, and Region C with the densest rim of the RMC.
However, Region D is found through the plots to have comparatively lower 
extinction (Fig. 4a), in agreement with results from 
the CO \& C$^{13}$O detections (Blitz \& Stark 1986; Williams et al. 1995).
The dust temperatures, however, are between 25 and 41~K (Fig. 4b), comparable to other regions
of cluster formation.

Star formation is evidenced by our study to be occurring in an extensive and regulated way across 
the south-east section of the RMC. This refines the picture introduced by Cox et al. (1990).
It is taking place within a selected spread but not all of the 
clumpy area of Region D (Fig. 4a), as is illustrated by the stellar distribution 
and the overplotted continuous lines in Fig.~5. 
Nevertheless, it follows a structured pattern that
first developed along the major axis of the RMC, in parallel to the Perseus Arm in the 
Galactic plane (Fig. 5). The pattern, resembling a 'tree' in structure, is related
to the gas filaments of the cloud (Williams et al. 1995), but does not
seem to break where the clumps or filaments do. This indicates the existence of
some underlining tracks.

A schematic model is thus invoked to interpret the widespread star and cluster 
formation activity in Regions C \& D of the RMC and, perhaps, the south-east half
of the complex as a whole. 
Large scale turbulence is suggested to originate from the spiral density wave 
of the Galaxy (Woodward 1976) or tidal force dissipation of the clouds in the Perseus arm. 
Additional turbulence is injected from internal sources, through 
a combination of feedback effects from the young massive stars in NGC 2244 and Region C, 
especially the violent outflow activity related to the formation of the proto-O stars
associated with AFGL 961 (Lada \& Gautier 1982).  The perturbations trigger 
the collapse of clumps, and star formation consequently occurs 
along tracks following the decay of the macroturbulence, tracing out a tree pattern.

In the frame of the proposed tree model, the rich massive cluster embedded in 
Region C is spatially coincident with the bulk of the CO emission 
of the complex (Blitz \& Thaddeus 1980; Blitz \& Stark 1986; Williams et al. 1995) 
and constitutes the root of the tree. It harbors the most massive protostars 
in the complex and perhaps also among those known in the Galaxy.  
Medium mass clusters and loose aggregates with stellar masses of up to $\sim$ 50 
M$\odot$ are developing down to the south-east boundary of the complex. These 
correspond to the branches.  
This scenario is supported by 
simulations of decaying supersonic turbulence, in which the decay
follows a period of energy pumping and compression (Mac Low \& Klessen 2004), 
but in this case it corresponds to wider or even galactic scales. 

This extensive study of the RMC has, for the first time, incorporated into the
scenario of sequential formation of OB clusters in GMCs the involvement of medium
to low mass stars which, in combination, reveal a more complex nature. We argue
that the history 
of cluster formation associated with the RMC, though in a sequential manner
both in space and time,
is better interpreted in terms of sequential collapse of dense clumps along the
major axis of the GMC. The collapsed regions outline tracks of the decay of macroturbulence rather
than having been exclusively triggered by the interference of former generations of
OB clusters of the Mon OB2 association.

{\flushleft \bf Acknowledgments~}

We are grateful to an anonymous referee for constructive comments 
which improved the scientific presentation of the paper. We
thank Prof. L. Blitz for going through the manuscript and providing very useful comments.
J. Z. Li acknowledges help from Dr. J. P. Williams for making available his
original CO \& C$^{13}$O data of the RMC.
This publication makes use of data products from the Two Micron All Sky Survey,
which is a joint project of the University of Massachusetts and the Infrared
Processing and Analysis Center/California Institute of Technology, funded by the
National Aeronautics and Space Administration and the National Science Foundation.
This work also makes use of the IRAS PSC and ISSA data. This project is supported
by SRF for ROCS, SEM. Finally, we acknowledge funding from the Particle Physics and
Astronomy Research Council, UK, and the Department of Culture, Arts and Leisure,
Northern Ireland.

\bibliographystyle{aa}

\begin{thebibliography}{}

\bibitem[]{331}Aharonian, F. A. et al. 2004, A\&A 417, 973
\bibitem[]{332}Bessel, M. S. \& Brett, J. M., 1988, PASP 100, 1134
\bibitem[]{333}Blaauw, A., 1964, ARA\&A 2, 213
\bibitem[]{334}Blitz, L. \& Stark, A. A., 1986, ApJ 300, L89
\bibitem[]{335}Blitz, L. \& Thaddeus, P., 1980, ApJ 241, 676
\bibitem[]{336}Block, D. L., Geballe, T. R. \& Dyson, J. E., 1993, A\&A 273, L41 
\bibitem[]{337}Carpenter, J. M., Snell, R. L., Schloerb, F. P. \& Strutskie, M. F. 1993, ApJ 407, 657
\bibitem[]{338}Celnik, W. E., 1985, A\&A 144, 171
\bibitem[]{339}Cox, P., Deharveng, L. \& Leene, A., 1990, A\&A 230, 181 
\bibitem[]{340}Davies, R. D., 1963, Observatory 83, 172
\bibitem[]{341}Dorland, H. \& Montmerle, T., 1987, A\&A 177, 243
\bibitem[]{342}Elmegreen, B. G., \& Lada, C. J., 1977, ApJ 214, 725
\bibitem[]{343}Hanson, M. M., Geballe, T. R., Conti, P. S. \& BLock, D. L., 1993, A\&A 273, L44
\bibitem[]{344}Kuchar, T. A. \& Bania, T. M., 1993, ApJ 414, 664
\bibitem[]{345}Lada, C. J., 1987, IAUS 115, 1
\bibitem[]{346}Lada, C. J. \& Gautier III, T. N., 1982, ApJ 261, 161
\bibitem[]{347}Lejeune, T. \& Schaerer, D., 2001, A\&A 366, 538
\bibitem[]{348}Li, J. Z., 2005, ApJ 625, in press
\bibitem[]{349}Li, J. Z. \& Smith, M. D., 2005a, A\&A 431, 925
\bibitem[]{350}Li, J. Z. \& Smith, M. D., 2005b, AJ 129, in press
\bibitem[]{351}Li, J. Z. \& Chen, P. S., 1996, Chinese Astro. \& Astroph. 20, 445
\bibitem[]{352}Mac Low, M.~\& Klessen, R.~S.\ 2004, Reviews of Modern Physics, 76, 125 
\bibitem[]{353}Meyer, M. R., Calvet, N., Hillenbrand, L. A., 1997, A\&A, 114, 288
\bibitem[]{354}Ogura, K. \& Ishida, K., 1981, PASJ 33, 149
\bibitem[]{355}Park, B.-G., \& Sung, H. 2002, AJ 123, 892
\bibitem[]{356}Phelps, R. L., 1994, in The Nature and Evolutionary Status of Herbig Ae/Be stars, ed. P. S. The, M. R. Perez \& P. J. van den Heuvel (San Francisco:ASP), 339
\bibitem[]{357}Phelps, R. L. \& Lada, E. A., 1997, ApJ 477, 176
\bibitem[]{358}Rieke, G. H. \& Lebofsky, M. J., 1985, ApJ 288, 618
\bibitem[]{359}Schneider, N., Stutzki, J., Winnewisser, G. et al., 1998, A\&A 338, 262
\bibitem[]{360}Shipman, R. F. \& Carey, S. T., 1996, ApJ 469, L131
\bibitem[]{361}Singh, K. P. \& Naranan, S., 1979, Ap\&SS 66, 191
\bibitem[]{362}Smith, M. G., 1973, ApJ 183, 111
\bibitem[]{363}Townsley, L. K., et al. 2003, ApJ 593, 874
\bibitem[]{364}Turner, D. G., 1976, ApJ 210, 65
\bibitem[]{365}Williams, J. P., Blitz, L., Stark, A. A., 1995, ApJ 451, 252
\bibitem[]{366}Woodward, P. R., 1976, ApJ 207, 484

\end{thebibliography}

\clearpage

\figcaption[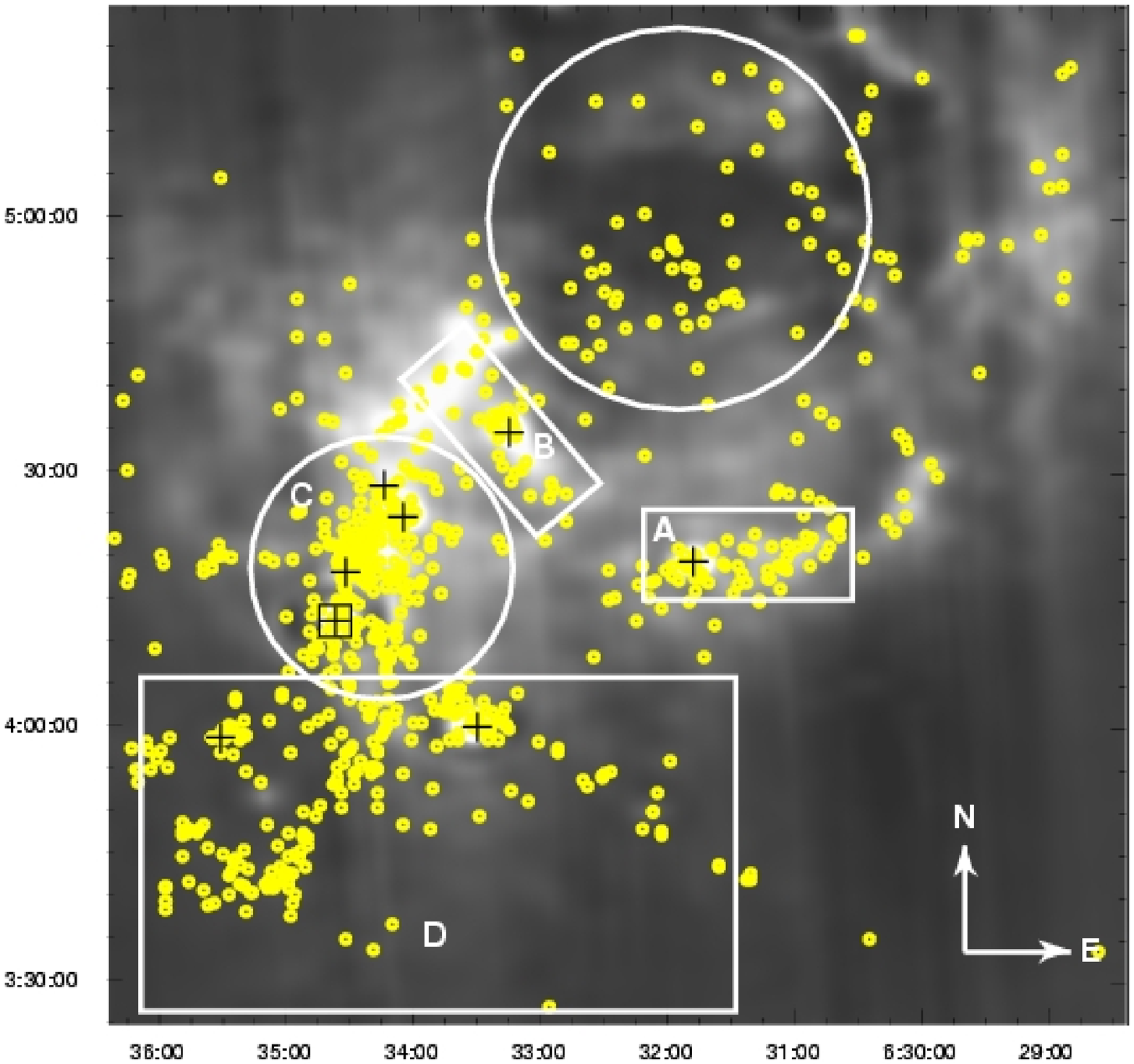]{The spatial distribution of the 2MASS sources with (H-Ks)$>$0.7 
in the RMC, with a field of 2$^\circ$\,$\times$\,2$^\circ$ centered at 
R.A.=06h32m24s, Dec.=+04d25m00s (J2000).
Well confined regions of clustered star formation are marked as A, B, C \& D. 
Contours of the IRAS HIRES 100 $\mu$m emission are superimposed onto the same
plot to further elucidate the spatial consistency of the embedded clusters with the
dense clumpy regions.  Positions of the 7 minor aggregations identified by 
Phelps \& Lada (1997) are plotted as pluses.  
The location of the candidate proto-O stars associated with AFGL\,961 is indicated by a solid square.
A dashed circle is drawn at the upper right of the plot indicating
the location of the young open cluster NGC\,2244 and its associated H{\small II} region in
expansion.}

\figcaption[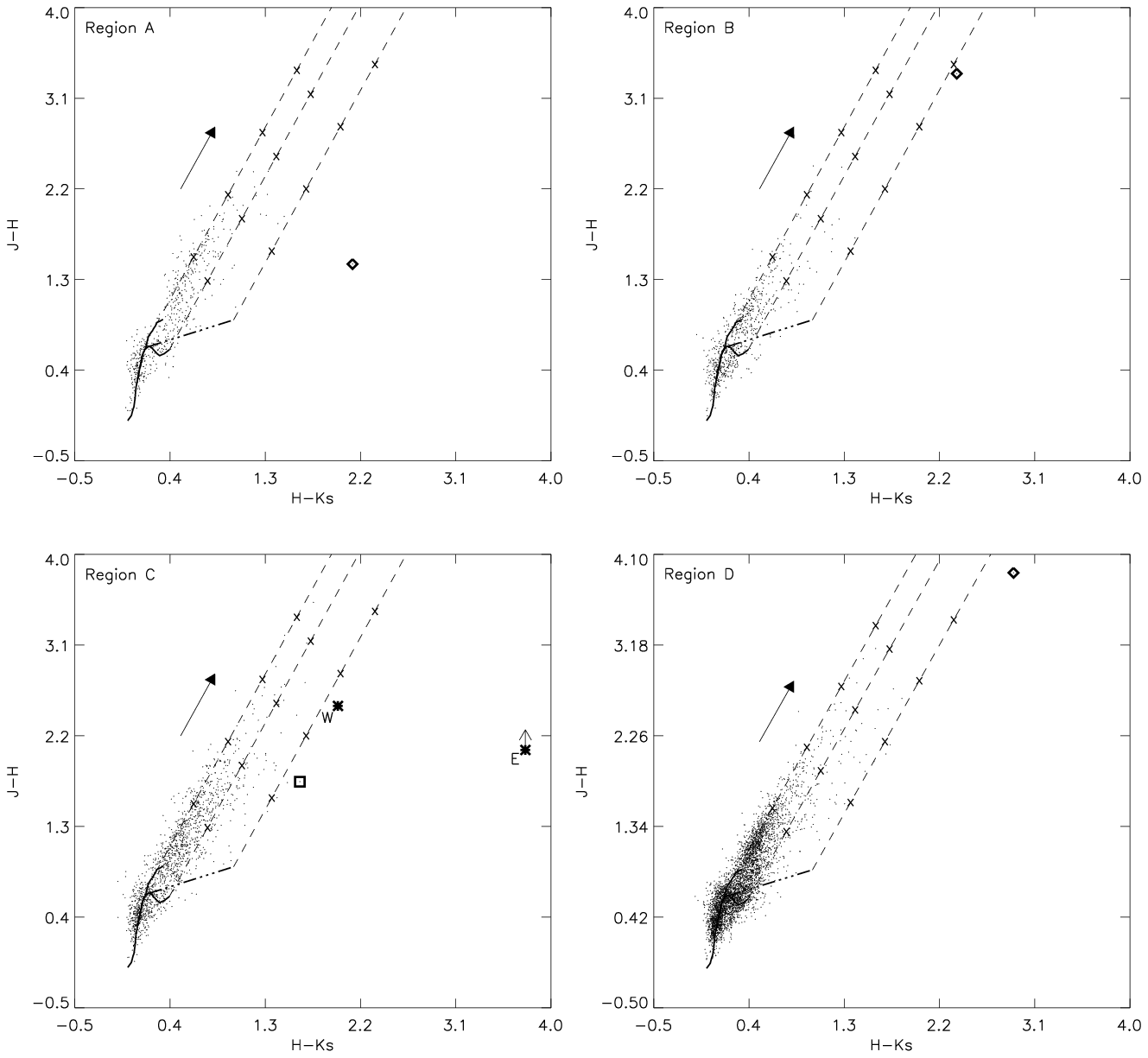]{Color-color diagrams of the outlined regions of on-going
cluster formation. The 2MASS sources that fit our selection criteria are plotted as dots.
Diamond symbols indicate the most extreme source in each corresponding 
region. Components of the conventional binary system AFGL 961 are shown with asterisks, 
the eastern component of which is further demarcated by an arrow pointing upwards 
indicating an upper limit detection in the J band and consequently a lower limit to 
J-H. The third component of the system is denoted as a square.
Solid lines represent the
loci of the main-sequence dwarfs and giant stars (Bessel \& Brett 1988). The
arrow in the upper left of the plot shows a reddening vector of Av = 5 mag
(Rieke \& Lebofsky 1985). The dot-dashed line indicates the locus of
dereddened T Tauri stars (Meyer et al. 1997). The dashed lines define the
reddening band for normal stars and T Tauri stars, and are drawn parallel
to the reddening vector, with crosses overplotted with an interval corresponding
to 5 mag of visual extinction.}

\figcaption[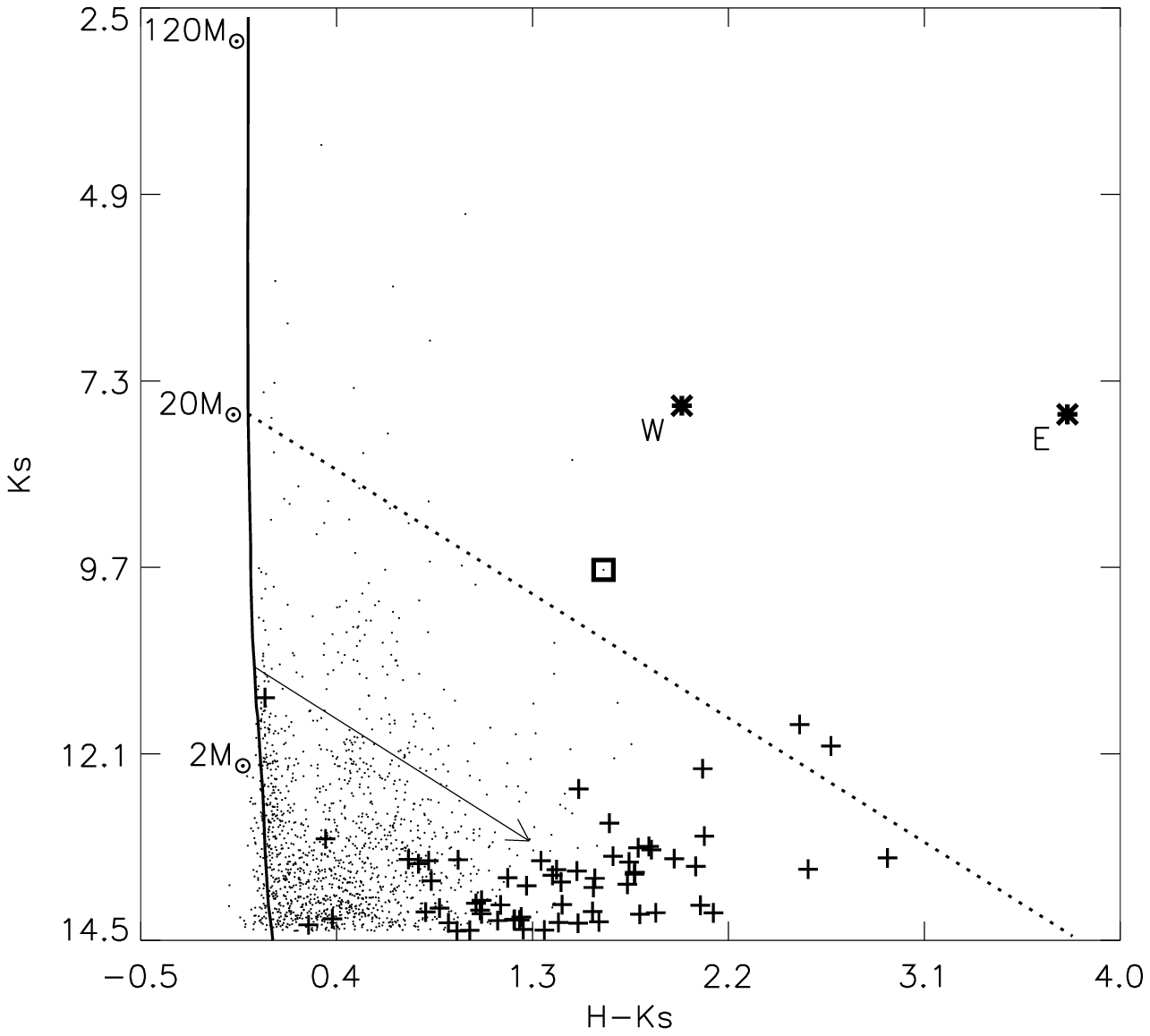]{A color-magnitude diagram of the Region C cluster and its
criteria missed candidate young stellar objects. Sample sources toward Region C
are plotted as dots and the criteria missed sources as pluses.
Both components of the massive binary AFGL 961, including the east component
that was overlooked by our source selection due to its upper limit
detection in the J band as discussed in \S~{2}, are indicated as asterisks. The third component
of the system, AFGL 961 II, is denoted as a square (see also Li \& Smith 2005b).
The main-sequence with masses between 0.8 and 120 M$\odot$
is plotted as a solid line (Lejeune \& Schaerer 2001). The slanted line with an
arrow at the tip denotes a reddening of Av = 20 mag of a A0 type dwarf.  The
dotted line indicates the reddening track of a B0 type star on the main
sequence.}

\figcaption[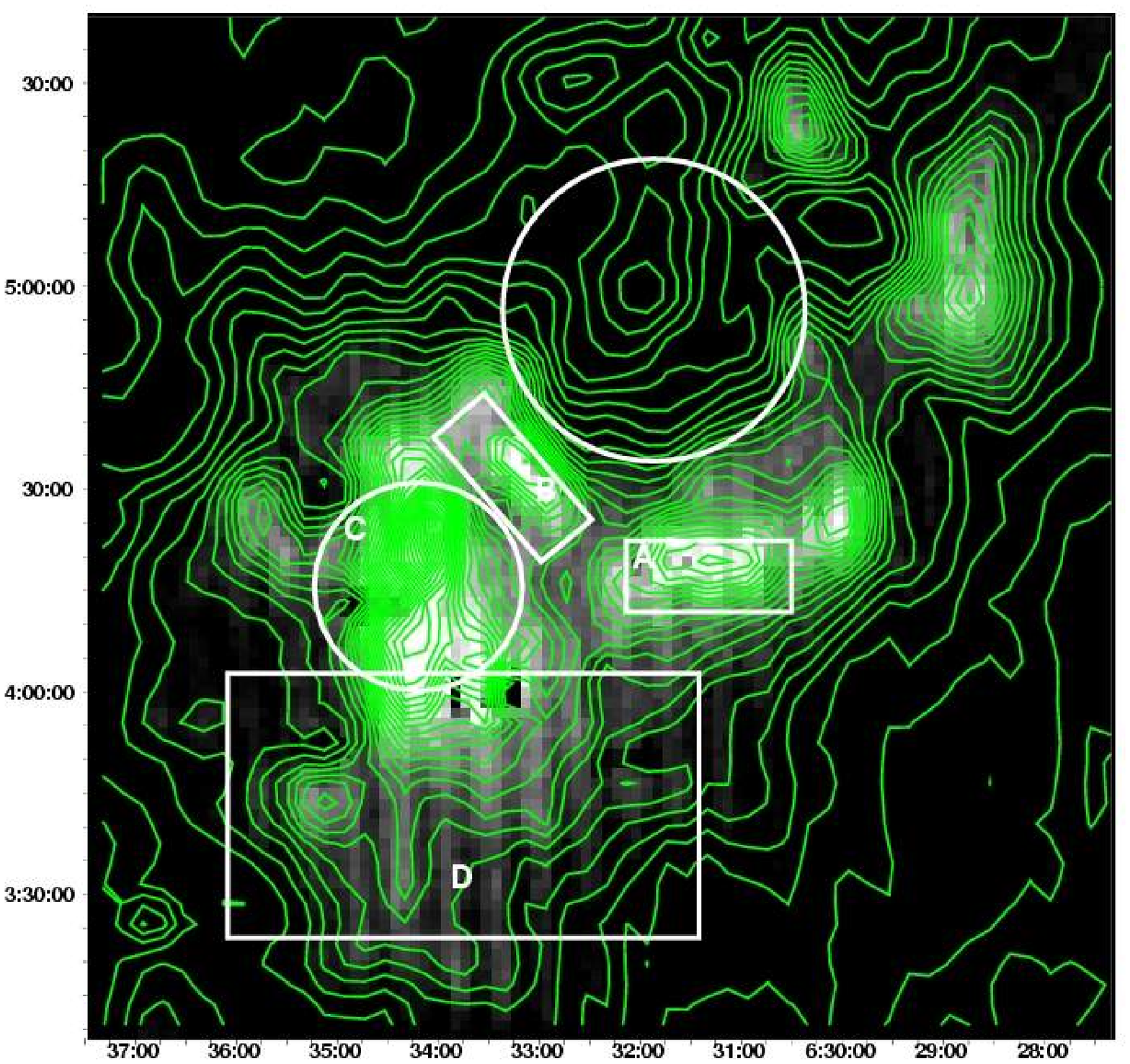,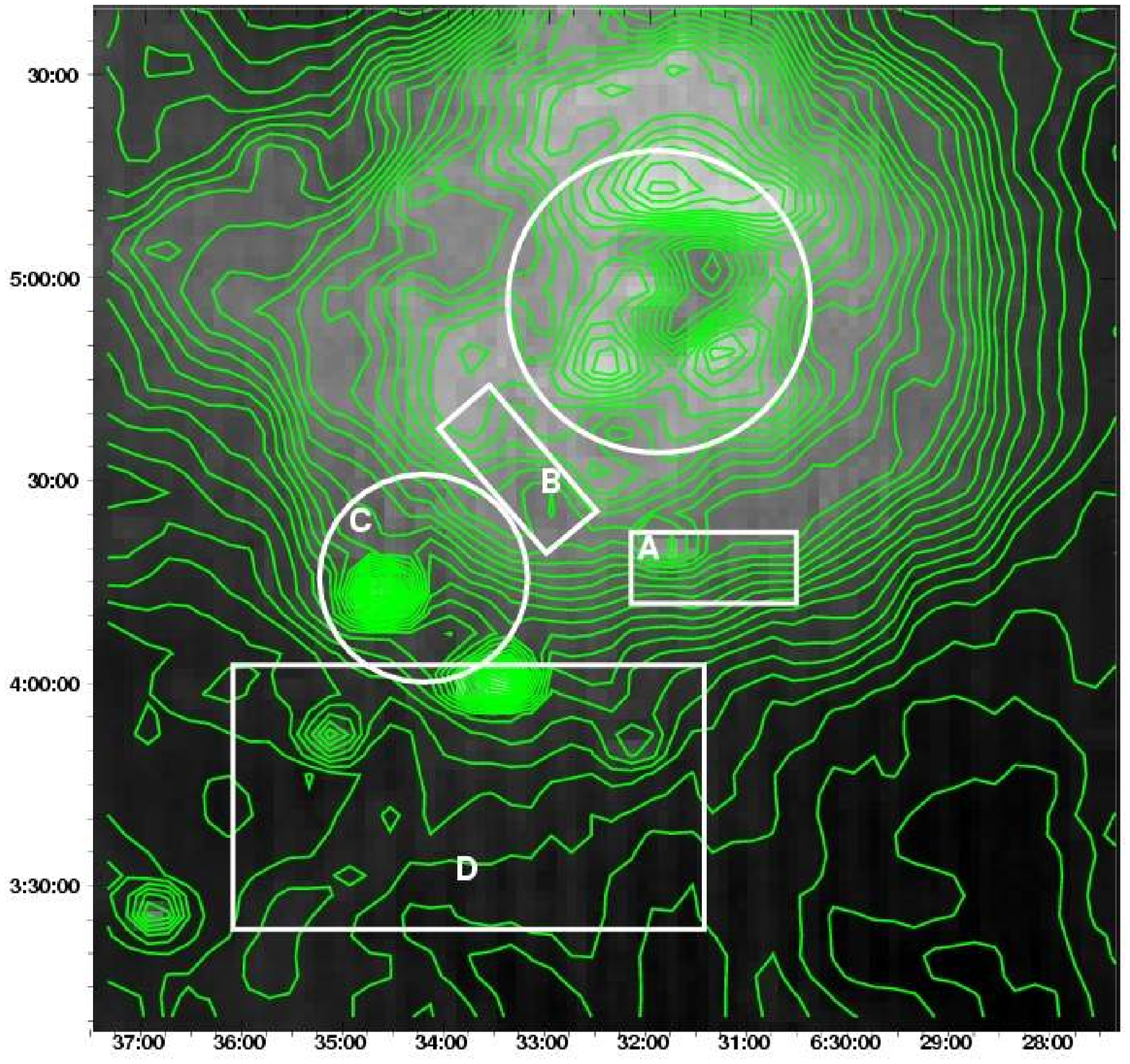]{a) Distribution of optical depth of the RMC
at 100$\mu$m derived from IRAS ISSA.  The overplotted contours 
are between 1.13 x 10$^{-4}$ and
1.16 x 10$^{-3}$, with a step of 1.2 x 10$^{-3}$. The fragmented interaction shells
of the Rosette Nebula with its surroundings is clearly seen as a circular structure,
where star formation is more or less emphatically going on. 
b) Distribution of effective temperature of cold dust in the RMC. The 
contours are drawn between 20.45\,K and 41.0\,K, with a step of 0.41\,K. Note the
distribution of condensed regions with higher cold dust temperature in the south-east of
the RMC. This, along with the regulated distribution of embedded sources in this
region, shows evidence that the dust temperature can be a key parameter governing 
the formation of stellar aggregates or clusters.}

\figcaption[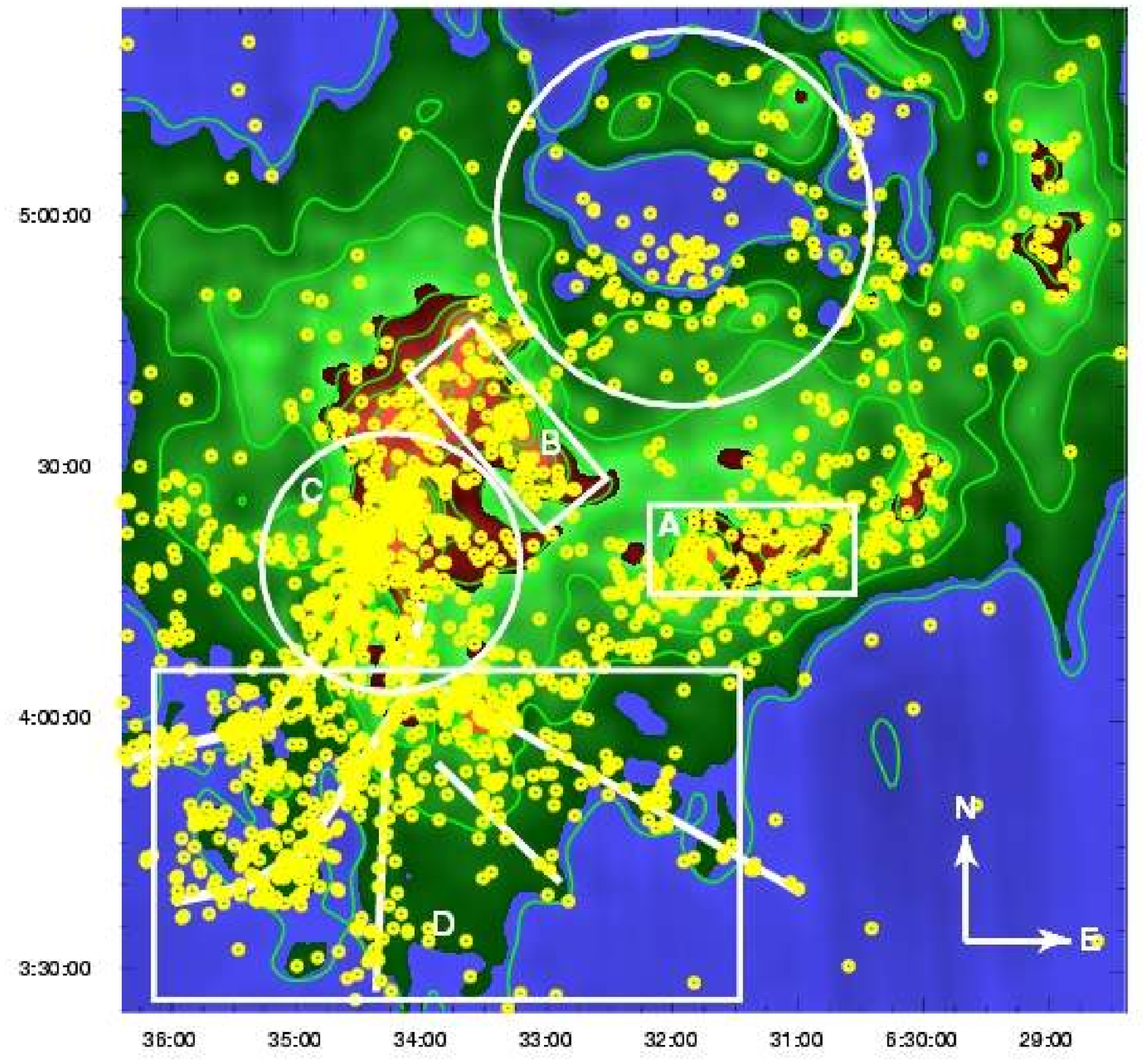]{The structured distribution of the sample sources with 
(H-Ks) ${>}$ 0.5 mag into a tree pattern, schematically illustrated by the
thick continuous lines.}

\newpage
\plotone{figure1.ps}
\newpage
\plotone{figure2.ps}
\newpage
\plotone{figure3.ps}
\newpage
\epsscale{0.6}
\plotone{figure4a.ps}
\plotone{figure4b.ps}
\newpage
\epsscale{1.0}
\plotone{figure5.ps}

\clearpage

\begin{deluxetable}{cccccccc}
\tabletypesize{\scriptsize}
\tablewidth{0pt}
\tablecaption{Near infrared clusters in the RMC} 
\startdata
\noalign{\smallskip}
\tableline
\noalign{\smallskip}
Sample &  Arbitrary center  &  Spatial scale &  Av  &  Weighted & Embedded & Evolutionary & Most massive\\
    &  ---------------------  &               &      &           &          &    & \\
Region &  RA (J2000)     Dec (J2000) &  (arcmin) & (mag) & mean (mag) & population & age (Myr) & object (M$\odot$)\\
\noalign{\smallskip}
\tableline
\noalign{\smallskip}
A & 06 31 28.20   04 19 06.6 & 27.90 x 15.78 & 0.2-15.0 & 7.0$\pm$0.5 & 41 & $\sim$ 1 & 20 \\
B & 06 33 10.80   04 34 40.8 & 7.20 x 19.40 & 0.5-19.5 & 2.5$\pm$0.5 & 47 & $\sim$ 1 & 20 \\
C & 06 34 18.48   04 19 51.6 & 32.00 x 32.00 & 0.5-21.5 & 4.0$\pm$0.5 & 123 & $<$ 1 & 130 \\
D & 06 33 49.50   04 30 04.5 & 74.25 x 38.33 & 0.5-24.5 & 4.0$\pm$0.5 & 251 & $<$ 1 & 50 \\
\noalign{\smallskip}
\enddata
\tablenotetext{*}{The sample region corresponding to each major cluster, an arbitrary central position, spatial extension of each sample area, visual extinction as estimated from the near infrared color-color diagrams, a weighted mean visual extinction toward each field, a lower limit of the embedded stellar population and the mass of the most massive component of each cluster.} 
\end{deluxetable}

\end{document}